# Optimizing run-length algorithm using octonary repetition tree

Kaveh Geyratmand Haghighi[1], Mir Kamal Mirnia*[2], Ahmad Habibizad Navin[3]

[1]Department of Computer ,East Azarbaijan Science and Research Branch,
Islamic Azad University,Tabriz, Iran
[1] Department of Computer Engineering, Tabriz Branch,
Islamic Azad University, Tabriz, Iran.

[2]Department of Computer Engineering,
University Of Tabriz, Tabriz, Iran

[3]Department of Computer ,East Azarbaijan Science and Research Branch,
Islamic Azad University,Tabriz, Iran

*Abstract*-**Compression is beneficial because it helps detract resource usage. It reduces data storage space as well as transmission traffic and improves web pages loading. Run-length coding (RLC) is a lossless data compression algorithm. Data are stored as a data value and counts. This is useful on data that contains many consecutive runs. This paper proposes a compression algorithm using octonary repetition tree (ORT), based on RLC. ORT is used to overcome the duplication problem in primary RLC algorithms, instead of using flag or codeword. It's the first method of run-length encoding which has the compression ratio greater than one in all tested cases. Experimental results, show average improvement of roughly 3 times, 3 times and 2 times in compression ratio field of study comparing to PRLC1, PRLC2, DF-RLC respectively. By using this approach of run-length encoding we can compress wider types of data, such as multimedia, document, executive files, etc.**

## I. INTRODUCTION

Digital image compression has become an important tool because image capturing devices, display units and social networks are widely used. When a raw image is stored as an array, usually the volume of file will be large and variable due to image resolution and color channel numbers. So, image compression may offer lower file size for storage and transmission. Usually there are two main approaches to compress data, such as lossy and lossless. Lossy image compression is irreversible, in the sense that it will lead missing details depending on the level of compression. Thus, this approach cannot be used in medical and satellite images and high level graphic design, since the original image should be remained with no changes. But, the original data can be reconstructed from the compressed data obtained from compression algorithm in lossless image compression.

Run-Length Coding (RLC) or run-length encoding [1], being part of JPEG [2], is a common lossless compression method which is used in International Telecommunication Union too. Run-length Coding is often used in other compression algorithms. In RLC sequence of bytes with the same value, encoded to number of repetition beside of byte value. For example, this sequence of values 338888888 is encoded to (3 2 8 7), where, 3 and 8 stands for the value of bytes, 2 and 7 shows the repetition correspondingly. The RLC is very efficient in encoding images when long runs of bytes appear with the same intensity value (eg., black and white images). In the other hand it is inefficient in encoding image with different values in sequence [5]. For example, 5 9 6 1 2 is a sequence of different bytes and will be encoded as (5 1 9 1 6 1 1 1 2 1) by RLC. In this case, adding number of repetition will increase data size (double in this case) and this is named as *duplication problem* in Run-Length Coding.

A new optimizing run-length algorithm is proposed here by using ORT. An octonary tree is added to point the sequence of bytes with the same value. The tree features is used to locate sequence of the same bytes with less overhead. Hence, in the proposed algorithm, there is no need to use extra bytes for determining the number of the same bytes, nor reserving a bit as a flag. The proposed approach has no duplication problem, and there is no limitation to encode number of bytes with the same value in a sequence. Also existence of an unused byte in image is unnecessary. To the best of our knowledge, the compression rate just in this method is greater than 1 in all tested images. Thus it is possible to use the compressing method for wider types of data, such as multimedia, documents, executive files, etc.

This paper is organized in five sections: Section 2 describes related works in RLC compression algorithm. Proposed method explains in section 3. Section 4 compares the new proposed method with recent works and section 5 concludes the paper and suggests future works.

## II. RELATED WORKS

As a solution to solve the duplication problem, [2], [5], [6] offered that RLC encoding occurs only when number of bytes with the same value is three or more and code other cases without using RLC. In this method, it is necessary to use unique flag bits for pointing byte intensity levels from number of occurrence in a stream of RLC encoded bytes. Two main techniques were suggested to implement the flag bits in RLC. The first technique uses a reserved and fixed codeword. This codeword treated as flag. The flag, locates a byte value with its count. In [6], flag is associated to a byte intensity value that does not exist. However, this technique is not acceptable if all of 256 bytes brought in the image. A similar approach that uses an unmatched codeword of one byte was applied in [7]. We named this technique as primary Run-Length 1 (PRLC1). Since a byte including 8-bits has range of [0, 255], count of consecutive bytes (with the same intensity value) that can be encoded with one run is at most 256. The main problem of PRLC1 is that, adding extra byte for separation run from byte value causes to duplication problem when different values in sequence appears. This approach could not Assurances that compression ratio is always greater than 1.

The second technique distinguishes count from value by defining MSB bit as a flag [8]. When MSB is set to 0, rest of bits (7-bits) indicates byte value and MSB=1, indicates following 7-bits is the count. This technique solves the duplication problem, but it is only acceptable when the image contains less than 128 unique values. Also, the count of consecutive bytes that can be encoded is at most 128. We named this technique as primary Run-Length 2 (PRLC2). Same as PRLC1, this method could not Assurances that compression ratio is always greater than 1.

Another approach in [10], named duplication free run-length coding (DF-RLC), uses rule-based generative coding to deriving codeword. This method solves duplication problem and the count of consecutive bytes that can be encoded is infinitive. This technique achieves better compression ratio than PRLC1 and PRLC2. But looks like other methods in this section, compression ratio is not always greater than 1.

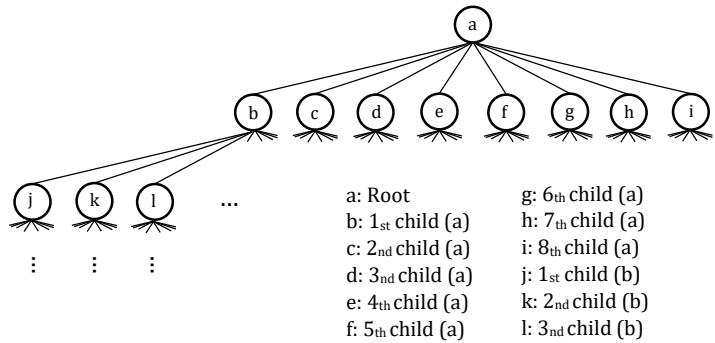

Figure 1: ORT structure

## III. PROPOSED METHOD

In the proposed method, a tree used to display duplicate bytes instead of using flag or codeword. This tree generates dynamically while reviewing the data. **Fig.1** has shown the generated tree structure Each node of this tree includes one byte (containing 8 bit). We named this tree "octonary repetition tree". This tree has 8 childes for each node, named k= {1,2, … ,8}.

The formula for calculating the array structure indices of the different relatives of a node are as follows:

*Parent(r) = ⌊(r−1)/8⌋ if r≠0.* (1)

$K_{th}$ *child(r) = 8r+k if 8r+k≤n.* (2)

The total count of nodes in the tree is *n* in (1), (2). The index of the node named as *r*, which is in the range of 0 to n−1. Each node in this tree has been ranged in [0-255] that represents a byte (containing 8-bits) in binary. The root of this tree has eight childes. As the lower layers has been reached, the data is divided into eight parts for each layer. This trend continues till leaves has reached. This method represents exact place of the duplicated byte. Each 8-bits in non-leaves, represents existence of equality of pair bytes in sub-branches. If there is no equality in sub-branches, calculating the array structure indices of the different relatives of a node algorithm prunes it. **Fig.2** has shown pruned tree for assumptive file including 64 bytes of data. Bytes with the same value are {(3,4,5),(7,8),(13,14,15),(51,52,53,54)}.

Table 1

| Position | 0 | 1 | 2 | 3 | 4 | 5 | 6 | 7 | 8 | 9 | 10 | 11 |
|---|---|---|---|---|---|---|---|---|---|---|---|---|
| Parent | - | 0 | 0 | 0 | 0 | 0 | 0 | 0 | 0 | 1 | 1 | 1 |
| $1_{st}$ Child | 1 | 9 | 17 | 25 | 33 | 41 | 49 | 57 | 65 | 73 | 81 | 89 |
| $2_{nd}$ Child | 2 | 10 | 18 | 26 | 34 | 42 | 50 | 58 | 66 | 74 | 82 | 90 |
| $3_{nd}$ Child | 3 | 11 | 19 | 27 | 35 | 43 | 51 | 59 | 67 | 75 | 83 | 91 |
| $4_{th}$ Child | 4 | 12 | 20 | 28 | 36 | 44 | 52 | 60 | 68 | 76 | 84 | 92 |
| $5_{th}$ Child | 5 | 13 | 21 | 29 | 37 | 45 | 53 | 61 | 69 | 77 | 85 | 93 |
| $6_{th}$ Child | 6 | 14 | 22 | 30 | 38 | 46 | 54 | 62 | 70 | 78 | 86 | 94 |
| $7_{th}$ Child | 7 | 15 | 23 | 31 | 39 | 47 | 55 | 63 | 71 | 79 | 87 | 95 |
| $8_{th}$ Child | 8 | 16 | 24 | 32 | 40 | 48 | 56 | 64 | 72 | 80 | 88 | 96 |

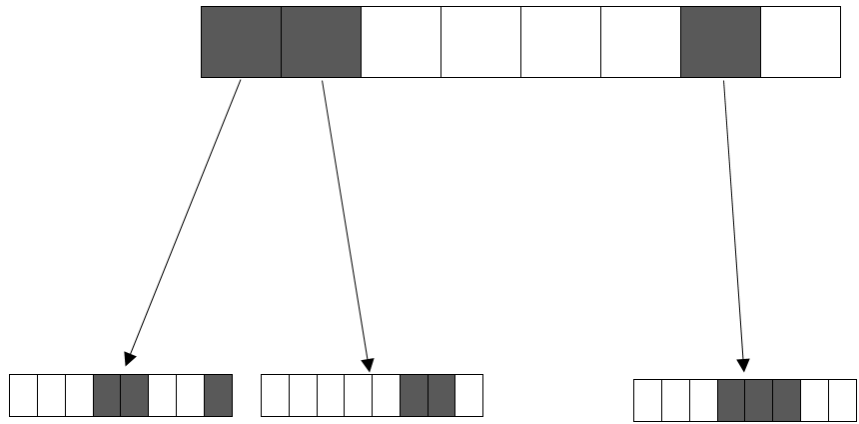

Figure 2. Pruned tree for assumptive file

Actually the non-leaf nodes are in charge of dividing the file. This division continues till the eight byte blocks has reached in data. Each 8-bits in leaves, shows equality of pair bytes in sequence. If bit value is set to 0, indicates pair bytes is not the same; when bit value is 1, means that pair bytes are the same. So, one byte in leaf of tree, shows similarity of 8 bytes in data. Also we defined limitation of similarity to construct leaves of tree. For example, limitation of 2, only allows to have leaves with at least 2 bits with value=1. Levels of the tree is determined by the size of the file. If $N$ is the total counts of bytes in file, then we calculate the number of levels as:

$$Level = \lceil log_8(N) \rceil \tag{3}$$

A fixed table defined to determine the depth of the tree and as it is fixed, it is embedded in the application and would not lead to further process. Proposed algorithm eliminates the same bytes in sequence while passes data and locates their place in tree. When algorithm reaches to end of file, attaches generated tree to end of it. This algorithm runs more than one time recursively to gain better results. In each run, we treat the output of algorithm as input and encode it again. In each iteration, comparison of bytes increases by one. **Fig.3** has shown these steps in flowchart. If $n$ is count of bytes and $k$ is number of runs. comparison of bytes at each run is as follow:

$$\sum_{k=0}^{10} byte[n] = \sum_{k=0}^{10} byte[n+k+1] \tag{4}$$

## IV. EXPERIMENTAL RESULTS

The ORT method is implemented in C# programing language software. Hardware used to run implemented program has an intel processor with 3.2 GHz frequency containing 8 thread and 8 GB ram with 1600 MHz frequency and SSD storage drive with reading and writing speed of 500 m/s. Windows 10 is also used as the operating system. and tested with 12 standard images [9]. Some of the test images is shown in **Fig4**.

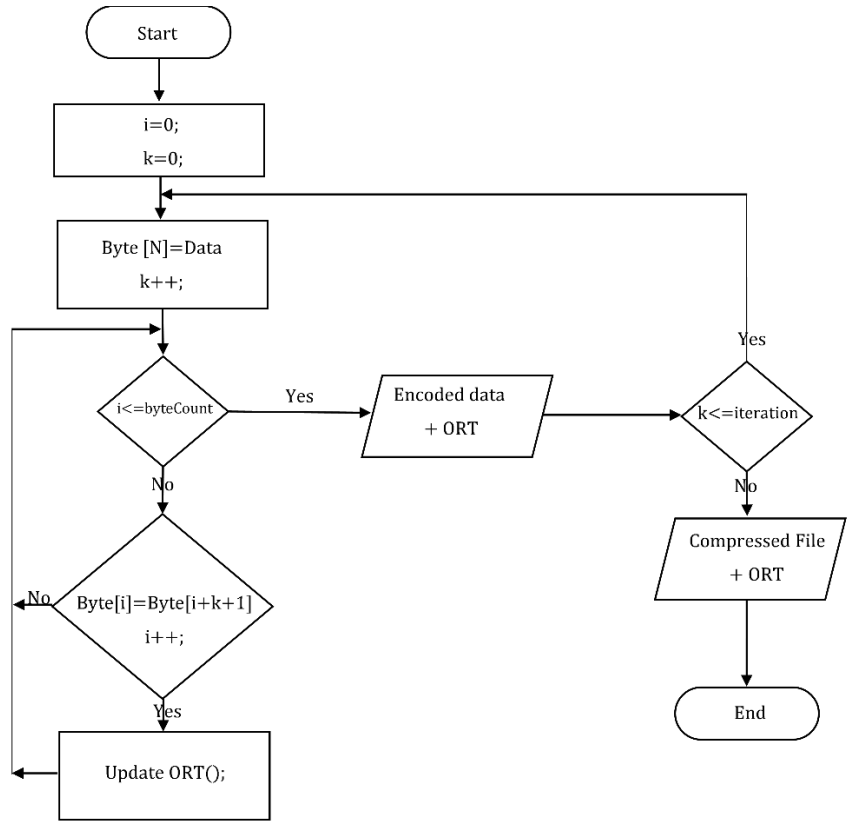

Figure 3. Flowchart of proposed algorithm

Compared with the methods used codeword for encoding, the proposed method in this article outs down the need for a codeword table. Compared with the methods used flag bit, this method has eliminated the limits of coding consecutive bytes (this problem exist in PRLC1, PRLC2). The proposed method in this article just as the PRLC1 does not have the duplication issue. The use of tree to display the duplicated bytes is an innovative method in this branch of encoding and has not be seen in any other article. One of the features of this tree is pruning any extra sub-branches to reach the least occupied byte. Generating individual tree (at the end of the file) also simplifies the uncompressing task. One of the important challenges in compressing is compression ratio. Data compression ratio (CR) is the ratio between the uncompressed and compressed size.

$$CR = \frac{Uncompressed\ Size}{compressed\ Size} \qquad (5)$$

The results of the proposed method shown in **Fig.5** and **Table 2**. Note that this results written after 10 times of running proposed algorithm recursively and limitation of the same bytes is >=3. In this table the comparison criterion is compression ratio. These results comprised with the PRLC1, PRLC2 and DF-RLC. If the compression ratio>1 it means the compression resulting data has gained free spaces after encoding. if compression ratio<1 indicates the increased size of data after encoding. Our approach has reached the CR>1 in any tested images. While in other methods, CR<1 has shown in some of cases.

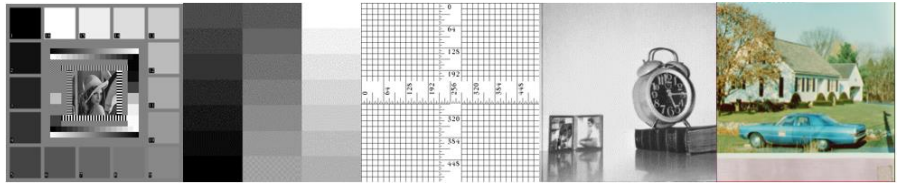

Figure 4. Test images

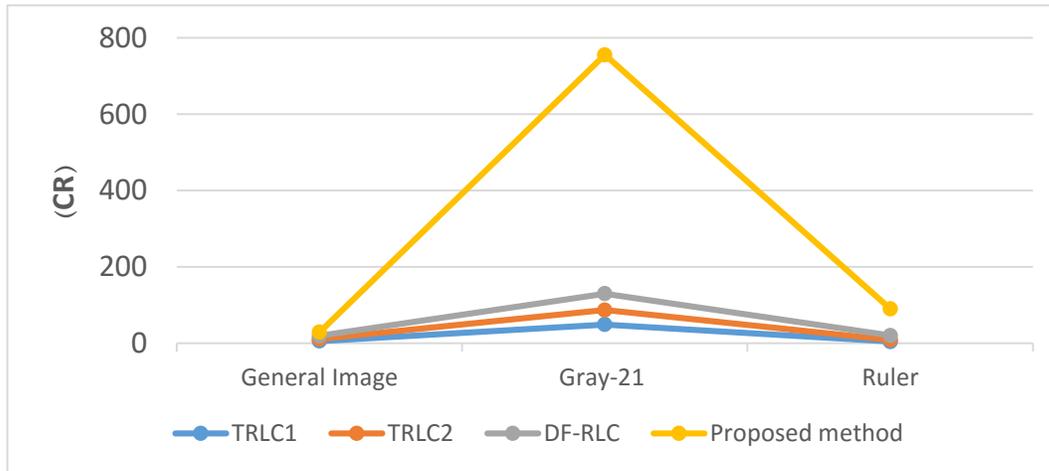

Figure 5. CR comparison of recent methods with proposed method

Table 2
Comparison of proposed algorithm compression ratio with other approaches

|    | Image         | Proposed method | DF-RLC | TRLC2 | TRLC1 |
|----|---------------|-----------------|--------|-------|-------|
| 1  | General Image | 9.467           | 8.3    | 5.4   | 5.5   |
| 2  | Gray-21       | 625.961         | 42.7   | 38.0  | 49.0  |
| 3  | Ruler         | 69.866          | 11.1   | 4.6   | 4.4   |
| 4  | Clock         | 1.113           | 1.1    | 0.9   | 0.9   |
| 5  | Tank          | 1.009           | 1.1    | 0.9   | 0.9   |
| 6  | APC           | 1.000           | 1.4    | 0.9   | 0.9   |
| 7  | Lenna         | 1.000           | 0.9    | 0.9   | 0.9   |
| 8  | Numbers       | 1.171           | 0.9    | 1.1   | 1.1   |
| 9  | Airplane      | 1.084           | 2.1    | 1.1   | 1.1   |
| 10 | F-16          | 1.021           | 1.0    | 0.9   | 0.9   |
| 11 | Boat          | 1.001           | 0.9    | 0.9   | 0.9   |
| 12 | House         | 1.050           | 0.9    | 0.9   | 0.9   |

New approach improved the compression ratio in all cases. The previous approaches which used the RLC for compression were limited to black and white or simple colored images due to their ineffectiveness. Hence these methods were limited to geographical and satellite and medical image. But the proposed approach in this article, is able to encode and compress other types of data such as multimedia files, applications, documents. In this method instead of using flag or codeword, a tree has been used to indicate the duplicated bytes. This tree is generated while passing the data and avoids reviewing the data repeatedly.

## V. CONCLUSION & FUTURE WORKS

In this paper run-length compression algorithm using ORT has been proposed. In this method, a tree used to display duplicate bytes instead of using flag or codeword. This method has eliminated the limits of coding consecutive bytes. In comparison to the traditional RLC approaches, this method overcomes to duplication problem. Our approach has reached the CR>1 in any tested cases and considered to the test results of compressed photos, improvements have been made in this regard. While in other methods, CR<1 has shown in some of images. It makes our algorithm suitable to use in compression of other types of data. The generated tree is pruned which leads to decreasing the number of bytes. We have a fixed table to determine the depth of the tree and since it is fixed it is embedded in the application and would not lead to further process. Parallelizing the algorithm by dividing the data and pipelining the recursive runs, speeds up compressing.

## References

[1] Bentley, J. L., et al. (1986). "A locally adaptive data compression scheme." Commun. ACM 29(4): 320-330.

[2] Acharya, T. (2006). "VLSI Algorithms and architectures for JPEG2000." Ubiquity 2006(September): 1-42.

[3] Togneri, R. and J. Christopher (2003). Fundamentals of information theory and coding design, CRC Press.

[4] Zhan, W. and A. El-Maleh (2009). A new collaborative scheme of test vector compression based on equal-run-length coding (ERLC). Proceedings of the 2009 13th International Conference on Computer Supported Cooperative Work in Design, IEEE Computer Society: 21-25.

[5] Luse, M. (1993). Bitmapped graphics programming in C++, Addison-Wesley.

[6] (2008). "Data Compression: The Complete Reference (by D. Salomon; 2007) [Book review]." IEEE Signal Processing Magazine 25(2): 147-149.

[7] Hughes, J. F., et al. (2014). Computer graphics: principles and practice, Pearson Education.

[8] Glassner, A. (1991). "Adaptive Run-Length Encoding." Graphics Gems II 2: 89.\

[9] "USC-SIPI image database." [Online]. Available http://sipi.usc.edu/services/database/

[10] Al-Wahaib, M. S. and K. Wong (2010). A Lossless Image Compression Algorithm Using Duplication Free Run-Length Coding. Network Applications Protocols and Services (NETAPPS), 2010 Second International Conference on.